\documentstyle[12pt,aaspp4]{article}
\begin{document}
\title{ An X-ray Flare Detected on the M8 dwarf VB 10 }

\author{ Thomas A. Fleming }
\affil{ Steward Observatory \\ 
University of Arizona \\ 
Tucson, AZ  85721 \\
taf@as.arizona.edu }
\author{ Mark S. Giampapa }
\affil{ National Solar Observatory \\
National Optical Astronomy Observatories\altaffilmark{1} \\
Tucson, AZ  85726 \\
giampapa@noao.edu }
\author{ J\"urgen H.M.M. Schmitt }
\affil{ Hamburg Sternwarte \\ 
Gojensbergsweg 112 \\
21029 Hamburg, Germany \\
st8h101@hs.uni-hamburg.de }

\altaffiltext{1}{The National Optical Astronomy Observatories are operated for
the National Science Foundation by the Association of Universities for 
Research in Astronomy.}
\begin{abstract}
   We have detected an X-ray flare on the very low mass star VB 10 (GL 752 B;
M8 V) using the ROSAT High Resolution Imager.  VB 10 is the latest type 
(lowest mass) main sequence star known to exhibit coronal activity.  
X-rays were detected from the
star during a single 1.1-ksec segment of an observation which lasted 22 ksec in
total.  The energy released by this flare is on the order of $10^{27}$  
ergs s$^{-1}$.  This is at least two orders of magnitude greater than the 
quiescent X-ray luminosity of VB 10, which has yet to be measured.  This X-ray
flare is very similar in nature to the far ultraviolet flare which was 
observed by Linsky et al. (1995) using the Goddard High Resolution Spectrograph
onboard the Hubble Space Telescope.  We discuss reasons for the extreme
difference between the flare and quiescent X-ray luminosities, including the 
possibility that VB 10 has no quiescent ($10^6$ K) coronal plasma at all.
\end{abstract}
\keywords{stars: X-rays, coronae, low mass }
\section{Introduction }
Stars at the very bottom of the main sequence have been a very popular 
topic of research during the last few years, including in studies of stellar
activity.
Fleming, Schmitt, \& Giampapa (1995) demonstrated that fully-convective M 
dwarfs, i.e. those with masses less than 0.3 M$_\odot$, exhibit high levels
of coronal activity, although at that time, virtually no X-ray detections 
had been made of stars less massive than VB 8 and LHS 3003 (both type M7.) 
The star VB 10 (a.k.a. GL 752 B, LHS 474; type M8)
appeared to have been marginally detected once by the Einstein High Resolution 
Imager and was catalogued by Barbera et al. (1993).  Then Linsky et al. (1995) 
reported the detection of a large amplitude 
flare on VB 10 in the far UV using the Goddard High Resolution Spectrograph.
More recently, Neuhauser \& Com\`oron (1998) have detected X-ray emission from
a very young brown dwarf in Chameleon.  This confirms that activity does exist 
for the very lowest mass stellar (or substellar) configurations.

We have used the ROSAT High Resolution Imager (HRI) to reobserve VB 10, which
was not detected by the ROSAT Position Sensitive Proportional Counter (PSPC)
during either the All-Sky Survey or a 7.3-ksec pointed observation (Fleming et 
al. 1993).  By obtaining a deeper exposure, we hoped to discover the nature
of the VB 10 corona, detect its quiescent X-ray luminosity, and compare its
level of coronal activity to that of the more massive M dwarfs.  We have 
succeded in detecting X-ray emission from VB 10, but only during a brief flare
similar to the one observed by Linsky et al. (1995) at ultraviolet wavelengths.

In this paper, we present the latest ROSAT data on VB 10.  Section 2
contains a description of the observations and data analysis.  In Section 3,
we discuss a hypothesis to account for the stark contrast between 
the flare and non-flare X-ray flux values, including the possibility of a 
total lack of $10^6$ K coronal plasma about VB 10.

\section{Observation}

The star VB 10 was observed during the period of 1997 October 27-31 for a total
of 21,992 sec with the ROSAT HRI (Zombeck et al. 1990).  
The total image is shown in
Fig. 1a.  The one obvious X-ray source seen in Fig. 1 is positionally 
coincident with the star Wolf 1055 (type M3.5 Ve),
which has a common proper motion with VB 10.   At the position of VB 10, which
is marked in Fig. 1a, there is only a marginal detection.  However, when
the data are separated into their individual Observation Intervals (OBIs),
we find that all of the photons in the VB 10 detection are contained within
one single OBI (Figs. 1b and 1c).

\begin{figure}
\caption{ROSAT HRI image of the field around VB 10:  a) the entire 22-ksec
exposure; b) a single 1.1-ksec OBI from UT 1997 October 29 3:05:24 to 3:25:43;
c) the remaining 20.9-ksec exposure.  The bright source seen in all three
images is Wolf 1055. All X-ray photons from VB 10 arrive in the single OBI in
Fig. 1b \label{fig1} }
\plotfiddle{f2.ps}{3.0in}{-90}{50}{50}{-210}{270}
\caption{Histogram of arrival times for the X-ray source photons from VB 10
\label{fig2} }
\end{figure}
This particular OBI, which began on UT 1997 October 29 at 3:05:24 and ended 
at 3:25:43, had an effective total exposure time of 1,138 sec.  Within an 
extraction radius of $12\arcsec$ around the optical position of VB 10,
11 photons were detected.  By examining the rest of the image, we determined 
that the density of background photons for this OBI was 0.0024 cts arcs$^{-2}$.
The solid angle of our extraction circle about VB 10 was 452.4 arcs$^2$,
which means that we would expect one photon in the extraction circle to 
be from the background.  Therefore, with 10 source photons, the mean count 
rate over the entire OBI is $8.9 (\pm 2.8) \times 10^{-3}$ cts s$^{-1}$ 
(this includes a deadtime and vignetting correction factor of 1.017.)

Of course, to get a better idea of the flare duration and flux, one needs
to look at the temporal distribution of the source photons throughout the 
observation.  In Fig 2, we show a histogram in arrival time for the 11 
photons contained within the source extraction radius.  Remember, we do not 
know which one is the background photon.  The histogram is binned in 3-minute
(180 sec) intervals over the nearly 20-minute observation.  One can see that
5 photons arrived in the first 3 minutes of the observation, 5 photons 
arrived during the last 8 minutes, with the remaining photon arriving in 
between.
The data are consistent with there being 
only one flare which is tailing off during our observation.  In this case, the
flare duration is at least 20 minutes and $2.8 \times 10^{-2}$ cts s$^{-1}$
(5 photons in 3 minutes) represents a lower limit to the peak count rate.


In order to get the X-ray luminosity, we adopt a conversion factor of 
$2.4 \times 10^{-11}$ ergs cm$^{-2}$ cnt$^{-1}$.  This comes from Table 10
of David et al. (1999), the ROSAT HRI Calibration Report (Cambridge: 
SAO)\footnote{$http://hea-www.harvard.edu/rosat/rsdc_www/hricalrep.html$}, 
for a Raymond-Smith spectrum of 0.5 keV
and negligible interstellar absorption.  This yields an apparent energy flux,
$f_X = 2.14 (\pm 0.68) \times 10^{-13}$ ergs cm$^{-2}$ s$^{-1}$ for 
the mean count rate.  At a distance of 5.74 pc,
this gives us a luminosity, $L_X = 8.4 (\pm 2.7) \times
10^{26}$ ergs s$^{-1}$.  Again, this is just a mean luminosity for the 
observation.  The peak flare luminosity would be at least $2.65 \times 10^{27}$
ergs s$^{-1}$.

We have also analyzed the remaining 20,854 sec of our HRI observation, in
which no X-ray source was detected at the position of VB 10.  Using the 
non-detection analysis software in MIDAS/EXSAS, we have calculated a $3\sigma$
(i.e. 99.7\% confidence) upper limit of $1.8 \times 10^{-4}$ cts s$^{-1}$.
This translates into $3\sigma$ upper limits on the apparent X-ray flux and
X-ray luminosity of $4.21 \times 10^{-15}$ ergs cm$^{-2}$ s$^{-1}$ and 
$1.7 \times 10^{25}$ ergs s$^{-1}$, respectively, for VB 10 outside of flare.
These are upper limits on any quiescent emission, if it exists.

All of the numbers presented in this section have been tabulated in Table 1.

\begin{deluxetable}{lclllr}
\footnotesize
\tablecaption{ X-ray Parameters for ROSAT HRI Observation of VB 10 
\label{tbl-1}}
\tablewidth{0pt}
\tablehead{
& \colhead{Exp Time} & \colhead{Count Rate} & \colhead{ f$_X$ } &
\colhead{ L$_X$ } & \colhead{ log (L$_X$/L$_{bol}$) } \\
& \colhead{ ks } & \colhead{ cts s$^{-1}$ } & \colhead{ $10^{-13}$ erg cm$^{-2}$ s$^{-1}$} 
& \colhead{ $10^{26}$ erg s$^{-1}$ } & }

\startdata

Flare (mean) & 1.14 & $0.0089 \pm 0.0028 $ & $2.1 \pm 0.7$ &
$8.4 \pm 2.7$ & $-3.3$ \nl
Flare (peak) & 0.18 & $> 0.028 $ & $> 6.7 $ &
$> 27 $ & $> -2.8$ \nl
Non-flare & 20.85 & $< 0.00018 $ & $< 0.042 $ &
$< 0.17 $ & $< -5.0$ \nl

\enddata
\end{deluxetable}

\section{Discussion}

This X-ray flare on VB 10 is reminiscent of the far-UV flare which was 
detected on VB 10 by Linsky et al. (1995).  These authors observed C II,IV
and Si IV emission lines only during the last five minutes of an hour-long
exposure taken with the GHRS onboard HST.  They concluded that the flare
which they had observed indicated increased magnetic heating rates for 
low-mass stars near the hydrogen-burning mass limit.  
 
In both the UV and ROSAT X-ray observations, quiescent (i.e. non-flare) 
emission 
was never detected from VB 10.  For the UV flare, the emission line fluxes 
were an order of magnitude greater than the upper limits placed on the 
non-flare emission by Linsky et al. (1995).  For this most recent X-ray flare,
the contrast is even greater.  The peak flare luminosity is {\it at least} 
more than two orders of magnitude greater than the non-flare value.

For ease of comparison to more massive M dwarfs, we will normalize L$_X$ 
by L$_{bol}$, which for VB 10 is $1.74 \times 10^{30}$ ergs s$^{-1}$ 
based on the absolute K magnitude measured by Leggett (1992) and the 
bolometric correction of Veeder (1974).
This gives us log (L$_X$/L$_{bol}) > -2.8$ for the peak of the flare and
log (L$_X$/L$_{bol}) < -5.0$ outside of flare.  Using the data of Fleming 
et al. (1995) for the late (later than M5), presumably fully-convective M
dwarfs within 7 pc of the Sun, we find that these star have values of 
log (L$_X$/L$_{bol}$) which are typically $-3.5$.  But during flares 
(e.g. AZ Cnc; Fleming et al. 1993), these stars can reach values of 
(L$_X$/L$_{bol}) = -3.0$ to $-2.5$.  Therefore, the magnitude of the flare on
VB 10 is completely consistent with that of flares on more massive M dwarfs.

In Fig. 3 we display a plot of normalized X-ray luminosity versus 
absolute visual magnitude for all known M dwarfs within 7 pc that are later 
than M5, based on data from a volume-limited survey by Fleming et al. 1995.
Inspection of Fig. 3 reveals the sharp contrast between the upper limit
to the non-flare X-ray emission in VB 10 and the X-ray emission levels of
other late dMe stars. In particular, the upper limit for VB 10 is
1-2 orders of magnitude less than that detected in the late dMe stars.  
It is comparable to the upper limit for GJ1002 in Fig. 3, a quiescent dM5.5 
star that does not exhibit H$\alpha$ emission nor any reported flare activity.
By contrast, VB 10 is a known flare star with (variable) H$\alpha$ line
emission in its spectrum.
While one cannot build
a theory based on one observation, we cannot help but speculate that our  
result for VB 10 does indeed reflect a decline 
in coronal heating
efficency near the H-burning mass limit.  But somehow these stars are still 
able to flare.

 We do not understand in detail the mechanisms that give rise
to energetic, transient outbursts identified as ``flares'' in the Sun and
late-type stars.  However, flares do appear to be the result of 
instabilities that return stressed systems toward configurations that
are characterized by lower potential energy (Rosner \& Vaiana 1978).
Since we do not have an X-ray spectrum, we cannot verify through 
modeling that the observed emission in VB 10 is consistent with 
the presence of loop-like magnetic structures.  However, the 
energetics of the event suggest the possible occurrence of a large 
volume of flare plasma, implying the presence of large-scale magnetic
structures in the atmosphere.  In particular, we can crudely estimate the 
plausible range of spatial scales that characterize the flare event in 
the following manner.

 In the absence of an actual energy spectrum (such as the pulse-height 
spectra that were produced by the now-defunct ROSAT PSPC), we 
assume some plausible flare plasma parameters.  Based on X-ray observations
of other flare events on M dwarfs, we adopt a flaring temperature of 
$T~\sim~$ 10$^{7}$ K.  We note that a large flare event recorded by 
the ROSAT PSPC on a late M dwarf star was well-described by a thermal
plasma model fit characterized by a temperature of 2 -- 4 $\times$ 10$^{7}$ K
(Sun et al. 1999).  Utilizing the XSPEC analysis package combined with 
the observed flux from the HRI and the adopted temperature yields a
differential emission measure at the flare maximum of EM $> 1.8 \times
10^{29}$ cm$^{-5}$.  The corresponding volume emission measure is
$$
VEM = 4 \pi R_{*}^2 EM > 1.1  \times 10^{50}~~cm^{-3}\,,
\eqno(1)
$$
where R$_{*}$ = 0.102 R$_{\odot}$ (Linsky et al. 1995).  Given this estimate
of volume emission measure along with electron densities in the range of
$n_e~\sim$ 10$^{10}$ cm$^{-3}$ to 10$^{11}$ cm$^{-3}$, we find 
characteristic linear 
dimensions for the flare in the range of 0.003 R$_{*}$ to 0.30 R$_{*}$.
Thus, the flaring plasma covers a large fraction of the stellar surface if
$n_e~\sim$ 10$^{10}$ cm$^{-3}$, but only a small fraction if 
$n_e~>$ 10$^{11}$ cm$^{-3}$.  We note that the distance traveled by a 
sound wave in a 10$^{7}$ K plasma in the 3 minute duration of the flare
maximum is about 0.7 R$_{*}$.  These estimates, while not conclusive,
are consistent with the likely occurrence of large-scale magnetic structures
associated with the flare event on VB 10.

 The enormous contrast between the X-ray flare luminosity and 
the upper limit to quiescent emission invites further consideration,
especially in view of the likely existence of significant surface magnetic 
flux and large magnetic structures in VB 10.  
We note that if VB10 is characterized by quiescent X-ray emission at a level
which is typical for the Sun at the maximum in its activity cycle, i.e.,
log($L_x/L_{bol})~\approx$ -6.3 (following Schmitt 1997), then we would still
not have had sufficient sensitivity to detect it.  The upper limit for 
non-flare X-ray emission in VB 10 is more comparable to the levels of emission
seen in earlier, more massive M dwarfs (non-dMe flare stars) which are 
themselves characterized by X-ray emission levels that are in excess of
solar, or log($L_x/L_{bol})~\sim$ -6 to -5 (following Fleming, Schmitt
\& Giampapa 1995).

 We thus cannot exclude the possibility that VB 10 does indeed have 
undetected, quiescent X-ray emission at the level of the Sun or that
of earlier,
quiescent dwarf M stars.  However, inspection of Fig. 3 suggests that the low
level of non-flare X-ray emission in VB 10 is unusual with respect to 
other late-type, active dMe flare stars. 
We will therefore briefly 
consider a hypothesis that may account for 
the extremely low, or even the possible absence of, steady, quiescent 
heating in this very cool dwarf. 

\begin{figure}
\plotfiddle{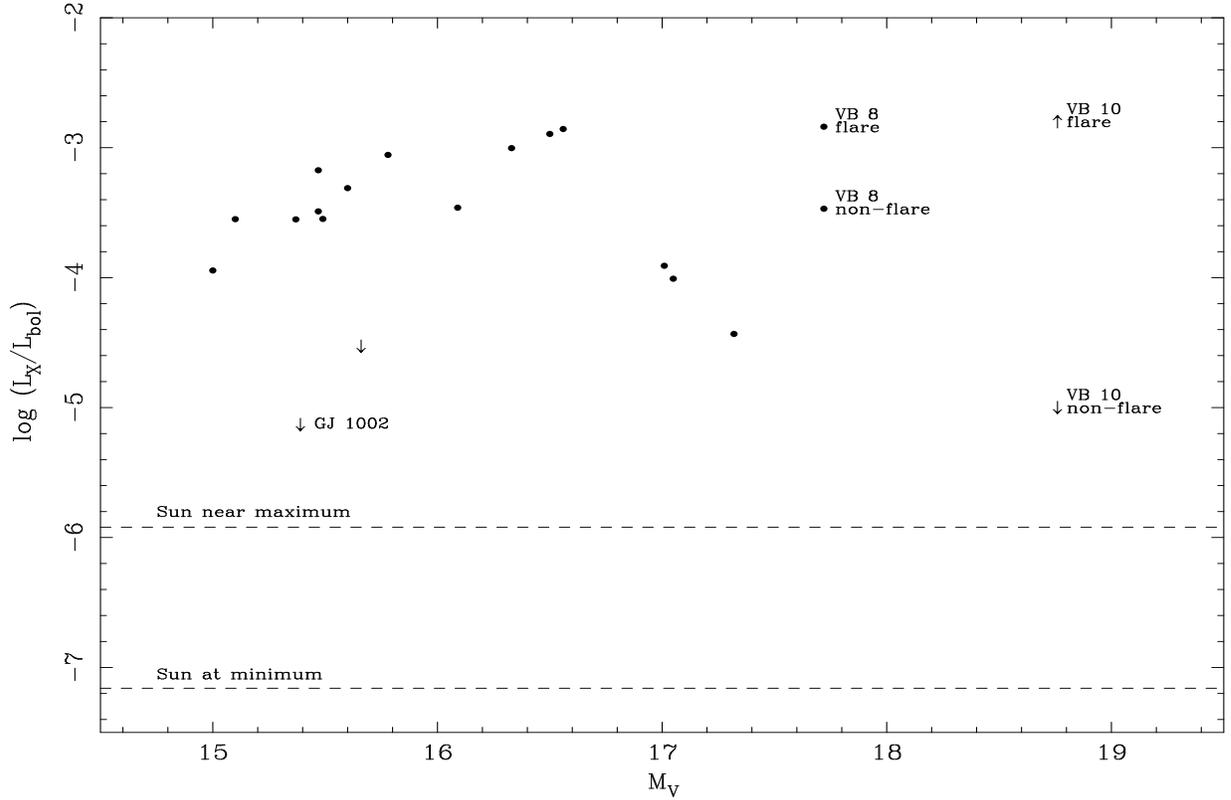}{2.0in}{-90}{70}{60}{-250}{350}
\caption{The normalized X-ray luminosity, log$(L_{x}/L_{bol})$, vs $M_{V}$
for M dwarfs later than M5.  Filled circles denote dMe stars.  Upper limits
are indicated by arrows.  Flare and non-flare values for certain objects
are shown. Data are taken from Fleming et al. (1993) and Giampapa et al. 
(1996).  Solar values are from Peres et al. (1999). \label{fig3} }

\end{figure}

 While there is no comprehensive theory for coronal 
heating---even in the case of the Sun---a common feature of current 
theories is that the origin of the nonradiative heating 
of the corona involves the interaction 
between turbulent convective motions in the photosphere and the 
footpoints of magnetic loops (Parker 1972, 1983a,b, 1986; van Ballegooijen
1986 and references therein).  
In this context, our observations of VB 10 may imply that at the 
photosphere there is simply
not enough in the way of random motions at the loop footpoints to jostle
the magnetic fields and, in turn, provide sufficient dynamical stresses 
to lead to detectable plasma heating to coronal ($T~\sim$ 10$^{6}$ K) 
temperatures.  For example,  a high magnetic field strength may suppress 
convective motions, analogous to the situation in sunspots which are 
also X-ray quiet.  We further note that the very cool and dense 
photosphere of VB 10 is dominated
in content by molecular hydrogen combined with neutral metals bound in 
molecules.  Thus, there are few ions to couple the magnetic field with the 
photospheric gas.  We estimate by extrapolation from 
M dwarf model photospheres (Mould 1976) to 
the effective temperature of VB 10 (T$_{eff}$ = 2600 K; Linsky et al. 
1995) that  
the ionization fraction, $\zeta$, in the dense, 
upper photosphere of VB 10 is $\zeta~\sim~$ 10$^{-7}$.  By contrast,
$\zeta~\sim~$ 10$^{-4}$ in the upper photosphere of the quiet Sun 
(Vernazza, Avrett, \& Loeser 1976).  In early M dwarfs we have 
that $\zeta~\sim~$ 10$^{-5}$ (following Mould 1976).     
Hence, for stars in the temperature-density regime of VB 10, the upper 
photospheres are characterized by ionization fractions that are 
2 orders of magnitude less than that of earlier M dwarfs and 
3 orders of magnitude below that of the Sun.
Consequently, the interaction between the field and the 
ambient gas occurs deeper in the photosphere where the ionization fraction 
is higher but where sufficient energy to produce significant coronal 
heating is unable to propagate outward.

 The consequences for coronal heating of the occurrence of magnetic 
structures in very cool dwarfs such as VB 10 is summarized in the following
argument due to F. Meyer (1999, private communication; see also 
Meyer \& Meyer-Hofmeister 1999).  Given that the 
gas pressure in the corona is typically negligible compared 
to the magnetic stresses, we must have that the magnetic field configuration,
or ``loop'', is force-free, or $\bf F = (J \times B)/{\rm c}$ = 
0.  Thus,  the electric currents must flow along the magnetic 
field lines. Since for steady magnetic phenomena the current density 
is divergence-free ($\bf \nabla \cdot J$ = 0), currents must also 
flow through the magnetic footpoints 
in the photosphere.  In the cool and dense photosphere of stars such as
VB 10, the electrical conductivity is so low that any current system 
rapidly decays.  From this it follows that the magnetic equilibrium must
be current-free everywhere.  
That is, $\bf J$ = 0 so that $\bf \nabla \times B$ = 0, and 
${\bf B = -\nabla}{\rm \Phi}$.  Given that the potential field is a minimum
energy configuration, any further build-up and storage of magnetic field
energy is excluded, implying no (or only a relatively low degree) of 
magnetic field-related heating of the atmosphere.  At the very least, the
decay rate of energy at the footpoints must be faster than any energy input 
derived from the interaction between motions in the upper photosphere
and the magnetic loops, effectively quenching any non-radiative heating
that might otherwise have occurred.

 Clearly, the above picture as outlined cannot explain flare events.
Instead, the transient or flare outbursts observed in these very cool 
stars must arise from more complex magnetic topologies where the storage of 
magnetic field energy occurs, but which do not have footpoints in the 
cool, dense underlying photosphere.

In summary, we have confirmed that the M8 dwarf VB 10 (Gl 752 B) does indeed 
emit X-rays.  It is the lowest mass star on the main sequence which is known
to do so.  This emission, however, appears to be only transient in nature.
The contrast between the flare X-ray flux and any possible non-flare 
(i.e., quiescent) X-ray emission is at least two orders of magnitude. 
We note that X-ray emission in VB 10 at solar levels or even at the 
levels seen in earlier dwarf M (non-dMe) stars would not be detected at 
our sensitivity limits.  In comparison to dMe stars later than M5, the 
upper limit to the non-flare X-ray emission in VB 10 is unusually low.

We have
considered a hypothesis that the low ionization fraction and, hence, low
conductivity, in the photosphere of VB 10 inhibits coronal 
heating to the point where quiescent, $10^6$ K coronal plasma may 
not even exist.  In this scenario, the transient, presumably $10^7$ K coronal 
plasma which
gives rise to the observed X-ray flare would then be associated with 
topologically complex
magnetic field structures that do not have footpoints in 
the cool photosphere which is, in turn, dominated by neutral atomic
and molecular species.  Should this scenario prove to be correct,
then there may indeed be a drop in coronal activity at the bottom of the 
main sequence:  not at the mass where stars become fully convective, as was
once suggested, but at the hydrogen-burning mass limit itself.

\acknowledgements
 We acknowledge insightful discussions with F. Meyer, B. Durney and
A. van Ballegooijen whose ideas materially contributed to this work.
TAF acknowledges support from NASA grant NAGW-3160.  MSG also acknowledges
support from NASA under the ROSAT Guest Observer program. The ROSAT project is 
supported by the German Bundesministerium f\"ur Forschung und Technologie 
(BMFT/DARA) and the Max Planck Gesellschaft.

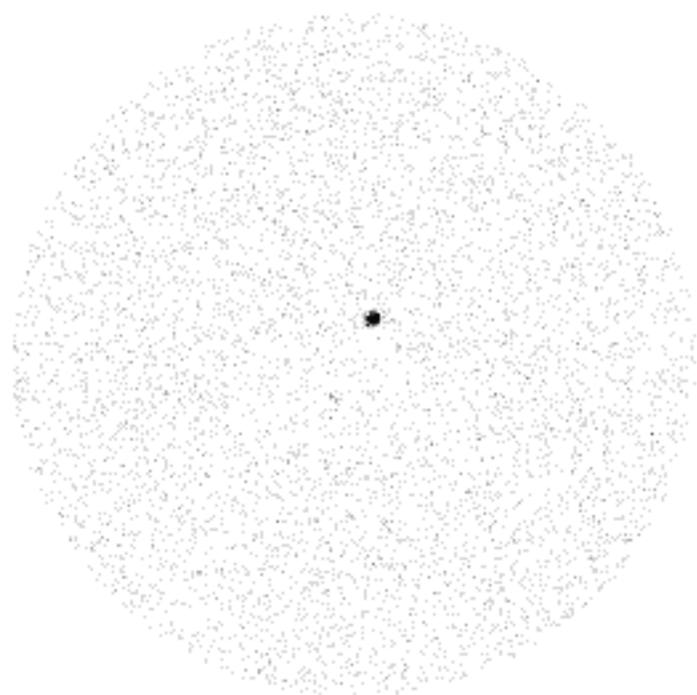
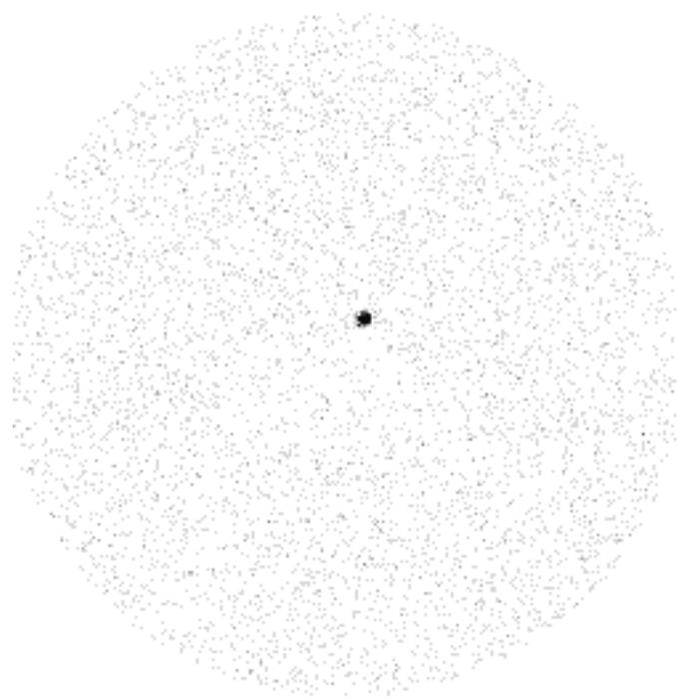
\end{document}